\documentclass[preprint,onecolumn,superscriptaddress,preprintnumbers,amsmath,amssymb]{revtex4-1}

\usepackage{xspace}
\usepackage{graphicx}
\usepackage{epstopdf}
\usepackage{color}

\newcommand{\rev}[1]{{\color{black}#1}}
\newcommand{\ups}{\upsilon}

\begin{document}

\title{Inverse proximity effect in semiconductor Majorana nanowires}
\author{Alexander~A. Kopasov}
\affiliation{Institute for Physics of Microstructures, Russian
Academy of Sciences, 603950 Nizhny Novgorod, GSP-105, Russia}
\email{kopasov@ipmras.ru}
\author{Ivan~M.~Khaymovich}
\affiliation{Max Planck Institute for the Physics of Complex Systems, D-01187 Dresden, Germany}
\affiliation{Institute for Physics of Microstructures, Russian
Academy of Sciences, 603950 Nizhny Novgorod, GSP-105, Russia}
\author{Alexander~S.~Mel'nikov}
\affiliation{Institute for Physics of Microstructures, Russian
Academy of Sciences, 603950 Nizhny Novgorod, GSP-105, Russia}
\affiliation{
Lobachevsky State University of Nizhny Novgorod, 23 Gagarina, 603950 Nizhny Novgorod, Russia}

\begin{abstract}
We study the influence of the inverse proximity effect on the superconductivity nucleation in hybrid structures consisting of the semiconducting nanowires placed in contact with a thin superconducting film and
discuss the resulting restrictions on the operation of Majorana-based devices.
A strong paramagnetic effect for electrons entering the semiconductor together
with spin-orbit coupling and van Hove singularities in the electronic density of states in the wire are responsible for the suppression
of superconducting correlations in the low field domain and for
the reentrant superconductivity at high magnetic fields in the topologically nontrivial regime.
The growth of the critical temperature in the latter case
continues up to the upper critical field destroying the pairing inside the superconducting film due to either
orbital or paramagnetic mechanism. The suppression of the homogeneous superconducting state
near the boundary between the topological and non-topological regimes
provides the conditions favorable for the Fulde-Ferrel-Larkin-Ovchinnikov instability.
\end{abstract}

\maketitle



\section{Introduction}
The transport phenomena in semiconducting wires with induced superconducting ordering and strong spin-orbit interaction
 are in the focus
of current experimental and theoretical research in field of nanophysics and quantum computing \cite{kitaev,Alicea1,Lutchyn,Oreg,nayak-rmp,alicea-nat,Aasen_comp,Alicea2,Elliott}. The interest to these systems is stimulated by the perspectives
of their use for design of topologically protected quantum bits. The key idea is based on the observation that for a certain range of parameters and rather strong applied magnetic field $H$ the
induced superconducting order parameter reveals so called p-wave symmetry realizing, thus, a model of Kitaev's chain \cite{kitaev}. The edges of such wires can host the subgap quasiparticle states which are considered as a realization of the Majorana particles in condensed matter systems
 \cite{Mourik2012,Chang2015,Higginbotham2015,Krogstrup2015,Albrecht2016,Leo}.

\begin{figure}
 \centering
 \includegraphics[width=0.45\textwidth]{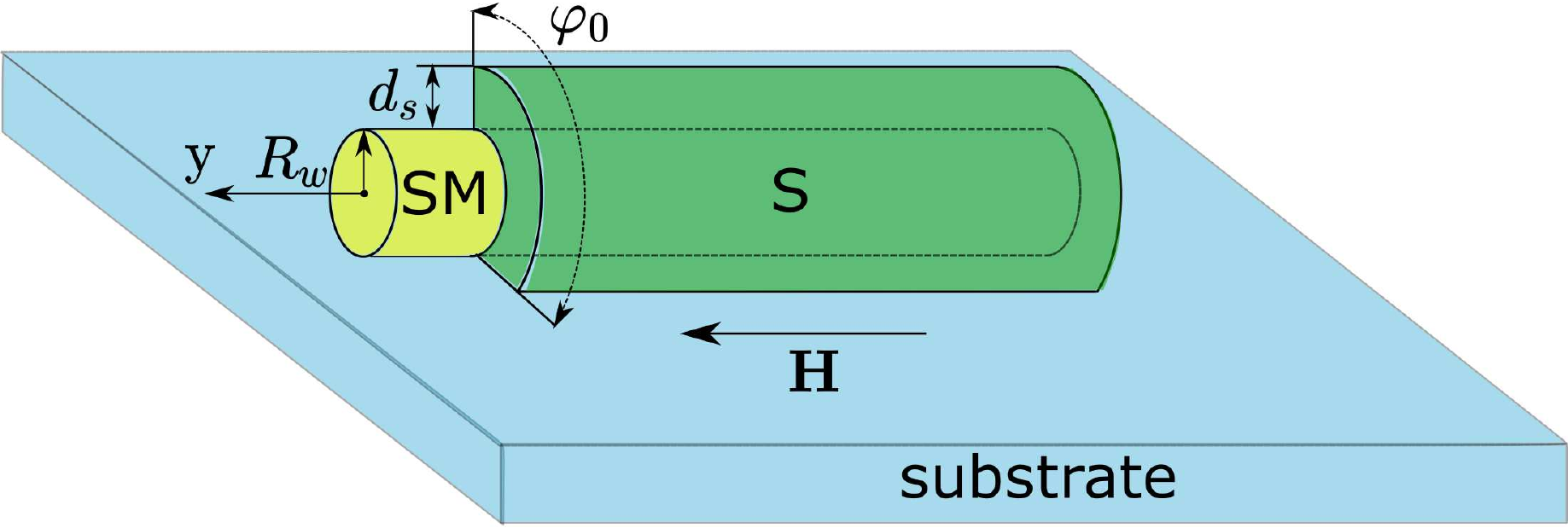}
 \caption{Schematic picture of the semiconducting wire (yellow) covered by the superconducting layer (green) placed on a substrate (light blue). $R_w$, $d_s$ and $\varphi_0$ show linear and azimuthal dimensions. The magnetic field $H$ is applied along the wire axis $Oy$ while the Rashba spin-orbit vector is perpendicular to the substrate (not shown).}
 \label{Fig:setup}
 \end{figure}

In most cases theoretical study of these Majorana wires is based on a simplified model of the superconducting correlations described by a
phenomenological gap potential inside the wire
 \cite{Lutchyn,Oreg}
placed in contact with a standard s-wave superconductor (see Fig.~\ref{Fig:setup}). Such model being useful in many cases for qualitative understanding of the induced superconductivity
is known to possess still a number of important shortcomings. An obvious way to overcome these shortcomings is to use the microscopic theory of the proximity effect
 \cite{McMillan,sau2010-2,Kopnin1,Kopnin2,Kopnin3,Stanescu2011}, i.e., Gor'kov equations.
The microscopic approach allows to get the effective gap operator analogous to the one used in the phenomenological model.
On top of that it gives
the gap dependence on the transparency of the interface between the wire and the s-wave superconductor and chemical potential via density of states (DOS).
 Another important point is that the exchange of electrons between the wire and superconductor can cause a so called inverse proximity effect, i.e., suppression of the gap function at the superconductor surface. For a rather thin superconducting shell covering the wire this gap suppression can result in the change of the superconducting critical temperature of the whole system. The analysis of this inverse proximity phenomena is important to find out the optimal range of parameters which allows to realize the switching between the topologically trivial and nontrivial states of the semiconducting wire used in various braiding protocols.

The goal of the present work is the self-consistent analysis of the critical temperature behavior in the wires with the induced superconducting ordering taking account of the inverse proximity effect. For this purpose we start from the full set of microscopic equations for the Green functions taking account of both scattering rates describing the quasiparticle transfer between the superconducting film and the wire \cite{McMillan}.
The first rate, $\gamma_s$,
characterizes the electron leakage from the wire to the superconductor and is responsible for the energy level broadening in the wire.
The second rate, $\gamma_w$, corresponds to the backward process. These rates are determined both by the probability of electron tunneling through the barrier at the superconductor/semiconductor (S/SM) interface and the corresponding densities of states. In particular, it is important that the rate $\gamma_w$ is proportional to the DOS in the SM nanowire resulting in its non-trivial energy dependence.
Indeed, considering, e.g., a single channel nanowire we get the DOS diverging as a
square root function of energy relative to the bottom of the conduction band. This van Hove singularity in the DOS should cause a strong energy dependence of the scattering rate $\gamma_w$ and, as a consequence, the superconducting critical temperature should depend on the position of the Fermi level with respect to the bottom of the one-dimensional conduction band in the SM wire. The influence of the van Hove singularity on
superconductivity should be also accompanied by the strengthening of the paramagnetic effect. Indeed, one can naturally expect that the scattering rate $\gamma_w$ could result in the additional effective Zeeman field induced in the superconductor due to the electron exchange with the SM wire.
Due to the divergence in the DOS together with the large $g$ - factor in the wire this induced Zeeman field can even exceed the usual Zeeman
field value. Under such conditions the field dependence of the critical temperature would have a minimum near the fields $H\sim |\mu_w|/g\beta$, where $\mu_w$
is the Fermi energy of the wire relative to the bottom of its conduction band at $H=0$ and $\beta$ is the Bohr magneton.
Strictly speaking, the spin-orbit interaction may cause the emergence of the third van Hove singularity below $-g\beta H/2$, but it appears only at rather large spin-orbit interaction strengths.
Note that for a vanishing induced superconducting gap $\Delta_{ind}$ this field separates the regimes with trivial and nontrivial topological properties of the system \cite{Lutchyn,Oreg,sau2010-2}. Further increase in the magnetic field is known to suppress the proximity effect since in the absence of the spin-orbit coupling the Fermi level crosses the only energy branch with a complete spin polarization along the magnetic field direction. The nonzero spin-orbit coupling destroys of course this spin polarization mixing different spin projections and resulting in a nonzero induced superconducting gap in the wire $\sim \alpha\Delta_{ind}/g\beta H$, where $\Delta_{ind}$ is the induced superconducting order parameter in the wire. Still even in the presence of the spin-orbit coupling the increasing magnetic field suppresses the induced superconductivity which definitely restores the superconducting order parameter in the S film. This reentrant superconductivity stimulated by the magnetic field can survive only up to the upper critical field associated with either orbital or intrinsic paramagnetic effect in the S shell.

The suppression of the superconducting order parameter near the line of transition between the topologically trivial and nontrivial phases can result in one more interesting phenomenon: similarly to the standard paramagnetic effect this suppression can cause the transition into the analog of the so-called Fulde-Ferrel-Larkin-Ovchinnikov (FFLO) state with the spatially modulated superconducting order parameter.

The rest of the paper is organized as follows. In the \rev{Section ``Basic equations''} we give the main equation of our model. The \rev{Section ``Results and Discussion''} is devoted to the description of the solution and the analysis of the phase diagrams. In the \rev{Conclusion Section} we summarize our results and the suggestions for the experiment.

\section{Basic equations}

Hereafter we consider a long 1D semiconducting wire partially covered by a thin superconducting shell with the thickness $d_s\ll \xi_s$, where $\xi_s$ is the superconducting coherence length. In the wire cross section the superconducting film covers the angular sector $\varphi_0$.
The model system is schematically shown in Fig.~\ref{Fig:setup}.
Hereafter we use the units with $k_B = \hbar = 1$, where $k_B$ is the Boltzmann constant, and $\hbar$ is the Planck constant.
The Hamiltonian of the system reads:
\begin{equation}\label{hamiltonian}
\mathcal{H} = \mathcal{H}_s+\mathcal{H}_w+\mathcal{H}_t \ ,
\end{equation}
\rev{with the} first term
\begin{gather}\label{hamiltonian_s}
\mathcal{H}_s = d_sR_w\int dy \ d\varphi \biggl[\psi_{\sigma}^{\dagger}(\mathbf{r})\varepsilon_s(\mathbf{r})\psi_{\sigma}(\mathbf{r})+
\Delta_s(\mathbf{r})\psi_{\uparrow}^{\dagger}(\mathbf{r})\psi_{\downarrow}^{\dagger}(\mathbf{r})+ \Delta_s^*(\mathbf{r})\psi_{\downarrow}(\mathbf{r})\psi_{\uparrow}(\mathbf{r})\biggl] \ ,
\end{gather}
\rev{describing the s-wave superconducting shell,
\begin{gather}
\mathcal{H}_w = S_{w}\int dy \ a_{\sigma}^{\dagger}(y)\Bigl[\varepsilon_w(y)-
i\alpha \hat{\sigma}_x \partial_y+h\hat{\sigma}_y\Bigl]_{\sigma \sigma'} a_{\sigma'}(y) \
\end{gather}
corresponds to the Hamiltonian of the nanowire, and
the tunnel Hamiltonian takes the form
\begin{gather}\label{hamiltonian_t}
\mathcal{H}_t = \sqrt{d_sR_wS_w} \int dy \ d\varphi [\psi^{\dagger}_{\sigma}(\varphi,y)\mathcal{T}(\varphi,y)a_{\sigma}(y)+
a^{\dagger}_{\sigma}(y)\mathcal{T}^{\dagger}(\varphi,y)\psi_{\sigma}(\varphi,y)] \ .
\end{gather}
Here $\sigma = \ \uparrow, \downarrow$ denotes spin degrees of freedom (summation over repeated spin indices is always assumed throughout the paper),
while $\hat{\sigma}_m$ ($m = x,y,z$) are the Pauli matrices in the spin space, $R_w$ is the radius and $S_{w}=\pi R_w^2$ is the cross section area of the wire, $\mathbf{r} =(R_w, \varphi,y)$, $\varphi$ is the polar angle in the plane perpendicular to the wire axis which changes in the interval $0<\varphi<\varphi_0$, $y$ denotes the coordinate along the wire,
$\varepsilon_s(\mathbf{r}) = -\nabla_{\mathbf{r}}^2/2m_s-\mu_s$ and $\varepsilon_w(y) = -\partial_y^2/2m_w-\mu_w$ stand for the quasiparticle kinetic energies in the shell and in the wire with respect to the corresponding chemical potentials $\mu_s$ and $\mu_w$, $m_s$ and $m_w$ are the effective masses of the electrons in the subsystems, $\Delta_s(\mathbf{r})$ is the superconducting order parameter, $\alpha$ is the spin-orbit coupling constant, $h = g\beta H/2$ is the Zeeman energy, and $H$ is the applied magnetic field.}

We consider the incoherent tunneling model, which does not conserve the momentum, e.g., due to the presence of the disorder at the interface. Thus, the ensemble average of the tunneling amplitudes has the form:
\begin{equation}\label{t_cor}
\overline{\mathcal{T}(\mathbf{r})\mathcal{T}(\mathbf{r}')} = t^2\ell_c\delta(y-y')\delta(\varphi-\varphi') \ ,
\end{equation}
where $\ell_c$ is the correlation length of the order of the atomic scale.
The tunneling is also assumed to be energy and spin independent and occurs locally in time and in space, i.e., from a point $\mathbf{r}$ on the superconducting shell into the point $y$ in the wire and back with the amplitude $\mathcal{T}(\mathbf{r})$.

It is important to note that here we do not consider the orbital effects in the superconducting shell.
This approximation of course imposes some restrictions on the value of magnetic fields under consideration which are nevertheless quite realistic for the
experiments aimed to the manipulation with Majorana states in such systems. It is the large $g$ factor in the SM wire which allows to have the magnetic field values affecting the electronic states in the wire and barely affecting the ones in the superconducting cover.
Note that omitting the orbital effects we cannot describe possible
Little-Parks effect arising in the wires fully covered by the S shell~\cite{Samokhvalov2009,Samokhvalov2017}.

\rev{Neglecting the order parameter inhomogeneity in the shell for $d_s\ll \xi_s$}, we derive the following system of Gor'kov equations written in the frequency-momentum representation (see the Appendix for the details of derivation)
\begin{eqnarray}
\label{gorkov_eqs_s}
 \left(i\omega_n-\varepsilon_s\check{\tau}_z+\check{\Delta}-\check{\Sigma}_s\right)\check{G}_s = \check{1} \ ,\\
 \label{gorkov_eqs_w}
 \left(i\omega_n-\varepsilon_w\check{\tau}_z-\alpha p_y\hat{\sigma}_x-h\hat{\sigma}_y-\check{\Sigma}_w\right)\check{G}_w = \check{1} \ ,
\end{eqnarray}
where $\omega_n = 2\pi T(n+1/2)$ is the Matsubara frequency, $T$~is the temperature, $\varepsilon_s$ is the single particle kinetic energy
relative to the chemical potential, $p_y$ is the momentum along the wire, $\check{\tau}_m$ ($m = x,y,z$) are the Pauli matrices acting in the Nambu space, $\check{\Delta} = (\hat{\Delta}\check{\tau}_++\hat{\Delta}^{\dagger}\check{\tau}_-)$, $\hat{\Delta} = \Delta_s(i\hat{\sigma}_y)$, $\Delta_s$ is the superconducting order parameter, which we assume to be constant in space and real-valued, $\check{\tau}_{\pm} = (\check{\tau}_x\pm i\check{\tau}_y)/2$, and $\varepsilon_w = p_y^2/2m_w-\mu_w$. The tunneling self-energy parts are given by the following expression:
\begin{equation}\label{self_energies}
\check{\Sigma}_{s(w)} = \Gamma_{w(s)}\check{\tau}_z\check{g}_{w(s)}\check{\tau}_z \ ,
\end{equation}
where $\rev{\Gamma_s = t^2\ell_c R_w m_s\varphi_0/2}$ and $\Gamma_w = t^2\ell_c/2\ups_0$. The functions $\check{g}_{s(w)}$ are the quasiclassical Green's functions:
\begin{eqnarray}
\label{quasiclassical_def_s}
\check{g}_s &= \frac{1}{\pi}\int d\varepsilon_s \ \check{G}_s(\omega_n,\varepsilon_s) \ , \\
\label{quasiclassical_def_w}
\check{g}_w &= \frac{\ups_0}{\pi} \int dp_y \ \check{G}_w(\omega_n,p_y) \ .
\end{eqnarray}
\rev{The precise definitions of the Green's functions $\check{G}_{w,s}$ of the wire and of the shell, respectively, together with the derivation of Eqs.~(\ref{gorkov_eqs_s}~-~\ref{gorkov_eqs_w}) are given in the Appendix.}

Note that we neglect here possible dependence of these quasiclassical Green's functions on the coordinate along the wire assuming, thus,
the limit of an infinitely long wire without edge effects.
The velocity $\ups_0$ is introduced just for the purpose of unification of dimensionality of the tunneling rates $\Gamma_w$ and $\Gamma_s$ and
does not appear in the product $\Gamma_w \check{g}_w $ which enters the measurable quantities.
One can choose this velocity, e.g., as $\ups_0=\sqrt{2\mu_w/m_w}$ so that the rate $\Gamma_w$ includes the divergent DOS in the 1D wire.

Tunneling rates for the quasiparticles in the shell $\Gamma_w$ and in the wire $\Gamma_s$ can be expressed in terms of the normal-state tunnel resistance $\mathcal{R}$ in the following manner \cite{Kopnin2}:
\begin{eqnarray}
\label{estimates_w}
\Gamma_w &= \frac{1}{4\pi\mathcal{R}S G_0\nu_s}\propto \mu_s\frac{\mathcal{R}_0}{\mathcal{R}}\frac{k_{Fw}}{k_{Fs}}\frac{1}{(k_{Fs}R_w)} \ ,\\
\label{estimates_s}
\Gamma_s &= \frac{1}{4\pi\mathcal{R}\ell_w G_0\nu_w}\propto \mu_w\frac{\mathcal{R}_0}{\mathcal{R}} \ ,
\end{eqnarray}
where $S = 2\pi R_w \ell_w$ is the contact area, $\ell_w$ is the wire length, $G_0 = e^2/\pi$ is the conductance quantum, $\nu_s = m_s/2\pi$ and $\nu_w=(2m_w/\mu_w)^{1/2}$ are the normal DOS in the shell and in the wire, respectively, $\mathcal{R}_0 = (NG_0)^{-1}$, $N = k_{Fw}\ell_w$, and $k_{Fs(w)}$ is the Fermi momentum in the shell (wire).~\rev{The expressions for the tunneling rates can be conveniently written through the numbers of transverse modes in the superconducting shell ($N_s\sim k_{F_s}^2d_sR_w\varphi_0$) and in the wire ($N_w$):
\begin{equation}\label{G_s}
\Gamma_s \sim t^2\ell_c\frac{N_s}{\ups_{F_s}} \ ,
\end{equation}
\begin{equation}\label{G_w}
\Gamma_w \sim t^2\ell_c\frac{N_w}{\ups_0} \ ,
\end{equation}
where $\ups_{F_s} = k_{F_s}/m_s$. Here we use the simplest generalization~\cite{McMillan} of the expression for $\Gamma_w$ for the case of an arbitrary number of transverse modes in the nanowire assuming also the value $1/\ups_0$ to be averaged over these modes. The resulting ratio of the tunneling rates takes the form:
\begin{equation}\label{tun_rates_ratio}
\frac{\Gamma_w}{\Gamma_s}\sim \frac{N_w}{N_s}\frac{\ups_{F_s}}{\ups_0} \ .
\end{equation}
Due to the growth of $N_s$ with the shell thinkness $d_s$ in the multi-mode regime of the superconductor this ratio may become rather small
weakening the inverse proximity effect (the details of experimental relevance are considered in the next section)
}
The Eqs.~(\ref{gorkov_eqs_s}~-~\ref{gorkov_eqs_w}) must be solved together with the self-consistency equation for the superconducting gap function:
\begin{equation}\label{SelfCons}
\Delta_s^* = \frac{\lambda \pi T}{2} \sum_{\omega_n} \text{Tr}\left[\left(\check{g}_s\right)_{21}\left(i\hat{\sigma}_y\right)\right] \ ,
\end{equation}
where $\lambda $ is the dimensionless pairing constant and the trace is taken over spin indices. The next section is devoted to the perturbative solution of the Gor'kov equations~(\ref{gorkov_eqs_s}~-~\ref{gorkov_eqs_w}) and the self-consistency equation~(\ref{SelfCons}) in the gap potential which allows to find the critical temperature of superconducting transition as a function of magnetic field and material parameters.

\section{Results and Discussion}
Considering the perturbation theory in the superconducting gap function
$\Delta_s$
it is natural to start with the equations for the normal Green's functions
\begin{eqnarray}
\label{normal_GF_G_s}
\left(i\omega_n - \varepsilon_s - \Gamma_w\hat{g}_w\right)\hat{G}_s = \hat{1} \ ,\\
\left(i\omega_n + \varepsilon_s - \Gamma_w\hat{\bar{g}}_w\right)\hat{\bar{G}}_s = \hat{1} \ ,\\
\left(i\omega_n - \varepsilon_w - \alpha p_y\hat{\sigma}_x - h\hat{\sigma}_y - \Gamma_s\hat{g}_s\right)\hat{G}_w = \hat{1} \ ,\\
\label{normal_GF_bar_G_w}
\left(i\omega_n + \varepsilon_w - \alpha p_y\hat{\sigma}_x - h\hat{\sigma}_y - \Gamma_s\hat{\bar{g}}_s\right)\hat{\bar{G}}_w = \hat{1} \ ,
\end{eqnarray}
which give us the zero order solution of the Gor'kov equations
\begin{eqnarray}
\label{normal_sols_gen_s}
\hat{G}_{s} &= \frac{i\omega_n-\varepsilon_s-\Gamma_wg_{w0}+\Gamma_w\mathbf{g}_w\hat{\boldsymbol{\sigma}}}{(i\omega_n - \varepsilon_s - \Gamma_w g_{w0})^2 - (\Gamma_w\mathbf{g}_w)^2} \ ,\\
\label{normal_sols_gen_w}
\hat{G}_{w} &= \frac{i\omega_n-\varepsilon_w-\Gamma_sg_{s0}+\mathbf{U}_w\hat{\boldsymbol{\sigma}}}{(i\omega_n - \varepsilon_w - \Gamma_s g_{s0})^2 - \mathbf{U}_w^2} \ .
\end{eqnarray}
Here $\mathbf{U}_w = \left[\left(\alpha p_y + \Gamma_sg_{sx}\right), \left(h + \Gamma_s g_{sy}\right), \Gamma_s g_{sz}\right]$ and the quasiclassical Green's functions are written in the spin form
\begin{equation}
\hat{g}_k = g_{k0} + \mathbf{g}_k\hat{\boldsymbol{\sigma}}\ ,
\end{equation}
with $k = s(w)$ for the shell (wire).
The solutions for spin matrix functions $\hat{\bar{G}}_k$ are given by the expressions~(\ref{normal_sols_gen_s}~-~\ref{normal_sols_gen_w}) with the replacement $\varepsilon_k \to -\varepsilon_k$ and $\hat{g}_k \to \hat{\bar g}_k $.

According to the definitions for the quasiclassical Green's functions~(\ref{quasiclassical_def_s}~-~\ref{quasiclassical_def_w}) and due to a specific spin structure of the Zeeman term and spin-orbit coupling term in the Eqs.~(\ref{normal_GF_G_s}~-~\ref{normal_GF_bar_G_w}), one can easily get that only $g_{k0}$ and $g_{ky}$ are nonzero. It is convenient to rewrite the normal Green's function in the wire as a sum of singular contributions $G_w^{p\pm}$:
\begin{gather}
\hat{G}_w = \frac{\left(\varepsilon_{so} + \varepsilon_0\right)G_w^{p+}-\left(\varepsilon_{so}-\varepsilon_0\right)G_w^{p-}}{2\varepsilon_0}-
\frac{\left(G_w^{p+}-G_w^{p-}\right)}{2\varepsilon_0}\left(\alpha p_y\hat{\sigma}_x + h\hat{\sigma}_y\right) \ ,
\end{gather}
\begin{gather}
\hat{\bar{G}}_w = \frac{-\left(\varepsilon_{so} - \varepsilon_0^*\right)\bar{G}_w^{p+}+\left(\varepsilon_{so}+\varepsilon_0^*\right)\bar{G}_w^{p-}}{2\varepsilon_0^*}-
\frac{\left(\bar{G}_w^{p+}-\bar{G}_w^{p-}\right)}{2\varepsilon_0^*}\left(\alpha p_y\hat{\sigma}_x + h\hat{\sigma}_y\right) \ ,
\end{gather}

where $G_w^{p\pm} = \left(i\bar{\omega}_n-\varepsilon_w + \varepsilon_{so}\pm \varepsilon_0\right)^{-1}$, $\bar{G}_w^{p\pm} = -\left(G_w^{p\mp}\right)^*$, $\varepsilon_0 = \sqrt{2\varepsilon_{so}\left(\mu_w + i\bar{\omega}_n\right)+\varepsilon_{so}^2 + h^2}$, $\varepsilon_{so} = m_w\alpha^2$, and $\bar{\omega}_n = \omega_n +\Gamma_s{\rm sign}(\omega_n)$. The equations for the anomalous Green's functions read:
\begin{eqnarray}
\label{anomalous_GF_s}
&\left(i\omega_n + \varepsilon_s - \Gamma_w\hat{\bar{g}}_w\right)\hat{F}_s^{\dagger}
=- \left(\hat{\Delta}^{\dagger}+\Gamma_w\hat{f}_w^{\dagger}\right)\hat{G}_s \ ,\\
\label{anomalous_GF_w}
&\left(i\omega_n + \varepsilon_w - \alpha p_y\hat{\sigma}_x - h\hat{\sigma}_y - \Gamma_s\hat{\bar{g}}_s\right)\hat{F}_w^{\dagger}
=- \Gamma_s\hat{f}_s^{\dagger}\hat{G}_w \
\end{eqnarray}
and give the solution for the anomalous Green's functions $\hat{F}_k^{\dagger}$ within the first-order perturbation theory in the superconducting gap
\begin{eqnarray}
\hat{F}_s^{\dagger} &= - \hat{\bar{G}}_s\left(\hat{\Delta}^{\dagger}+\Gamma_w\hat{f}_w^{\dagger}\right)\hat{G}_s \ ,\\
\hat{F}_w^{\dagger} &= - \hat{\bar{G}}_w\Gamma_s\hat{f}_s^{\dagger}\hat{G}_w \ .
\end{eqnarray}

Introducing a general presentation for the components of the quasiclassical anomalous Green's functions
\begin{gather}
\hat{f}_k^{\dagger} = -i\hat{\sigma}_y\left(f_{k0}^{\dagger} + \mathbf{f}_k^{\dagger}\hat{\boldsymbol{\sigma}}\right),
\end{gather}
we get the following set of equations for them 
${f_{sx}^{\dagger} =f_{sz}^{\dagger} =0}$
\begin{widetext}
\begin{equation}\label{anomalous_system}
\begin{pmatrix}-1+\gamma\left[I_{sy}I_{wy} + I_{s0}(I_{w0}-I_{wx})\right]& \gamma\left[I_{s0}I_{wy}+I_{sy}(I_{w0}+I_{wx})\right] \\ \gamma\left[I_{s0}I_{wy}+I_{sy}\left(I_{w0}-I_{wx}\right)\right]&-1 + \gamma\left[I_{sy}I_{wy}+I_{s0}\left(I_{w0}+I_{wx}\right)\right]\end{pmatrix}\begin{pmatrix}f_{s0}^{\dagger}\\ f_{sy}^{\dagger}\end{pmatrix} = \Delta_s\begin{pmatrix}I_{s0}\\ I_{sy}\end{pmatrix} \ .
\end{equation}
The solutions of the Eqs.~(\ref{anomalous_system}) take the form
\begin{eqnarray}
\label{anom_f_s0}
f_{s0}^{\dagger} &= -\Delta_s\frac{\left[I_{s0}-\gamma\left(I_{w0}+I_{wx}\right)\left(I_{s0}^2-I_{sy}^2\right)\right]}{\left[1-2\gamma\left(I_{s0}I_{w0}+I_{sy}I_{wy}\right)+\gamma^2\left(I_{s0}^2-I_{sy}^2\right)\left(I_{w0}^2-I_{wx}^2-I_{wy}^2\right)\right]} \ ,\\
\label{anom_f_sy}
f_{sy}^{\dagger} &= -\Delta_s\frac{\left[I_{sy}+\gamma I_{wy}\left(I_{s0}^2-I_{sy}^2\right)\right]}{\left[1-2\gamma\left(I_{s0}I_{w0}+I_{sy}I_{wy}\right)+\gamma^2\left(I_{s0}^2-I_{sy}^2\right)\left(I_{w0}^2-I_{wx}^2-I_{wy}^2\right)\right]} \ ,
\end{eqnarray}
\end{widetext}
Here and further we use the following notations
\begin{eqnarray}
\label{I_w_def}
&I_{\nu\eta}^w = \int \frac{dp_y}{\pi} \bar{G}_w^{p\nu}G_w^{p\eta} = \frac{\bar{g}_w^{p\nu}+g_w^{p\eta}}{2i\bar{\omega}_n+\nu\varepsilon_0^* + \eta\varepsilon_0} \ ,\\
\label{I_w_def_x}
&I_{\nu\eta}^{wx} = \int \frac{dp_y}{\pi}\frac{p_y^2}{2m_w}\bar{G}^{p\nu}_wG_w^{p\eta} = \frac{\Bigl[\bar{g}_w^{p\nu}\left(-i\bar{\omega}_n - \nu\varepsilon_0^* + \mu_w + \varepsilon_{so}\right) +
g_w^{p\eta}\left(i\bar{\omega}_n + \eta\varepsilon_0 + \mu_w + \varepsilon_{so}\right) \Bigl]}{(2i\bar{\omega}_n+\eta\varepsilon_0+\nu\varepsilon_0^*)}
\end{eqnarray}

\begin{eqnarray}
I_{wx} &=& \sum_{\nu,\eta = \pm 1}\nu\eta \frac{\epsilon_{so}I^{wx}_{\nu\eta}}{2|\varepsilon_0|^2} \ ,\\
I_{w0} &=& \sum_{\nu,\eta = \pm 1}\Bigl[
\left(\varepsilon_0^*-\nu \varepsilon_{so}\right)\left(\varepsilon_0+\eta \varepsilon_{so}\right)+
h^2\nu\eta\Bigr]\frac{I^w_{\nu \eta}}{4|\varepsilon_0|^2} \ ,\\
I_{wy} &=& \sum_{\nu,\eta = \pm 1}\left(\eta\varepsilon_0^*+\nu\varepsilon_0\right)\frac{(-h) I^w_{\nu \eta}}{4|\varepsilon_0|^2} \ .
\end{eqnarray}
Here and further $\nu,\eta =\pm 1$.

The expressions for the integrals involving the products of the normal Green's functions in the shell can be written as follows:
\begin{eqnarray}
\label{I_s0_def}
I_{s0} &=& I_{s+}+I_{s-} \ ,\\
I_{sy} &=& I_{s+}-I_{s-} \ ,\\
\label{I_seta_def}
I_{s\eta} &=& \int \frac{d\varepsilon_s}{(2\pi)}\bar{G}_{s\eta}G_{s\eta} = \frac{1}{2}\frac{\left(\bar{g}_{s\eta}+g_{s\eta}\right)}{\left[2i\omega_n - \Gamma_w\left(\bar{g}_{w\eta}+g_{w\eta}\right)\right]} \ .
\end{eqnarray}
In the definitions~(\ref{I_w_def}~-~\ref{I_w_def_x}) and~(\ref{I_s0_def}~-~\ref{I_seta_def}) we have introduced the following functions:
\begin{eqnarray}
\bar{g}_{s\eta} &=& g_{s\eta} = g_{s0} = -i\,{\rm sign}(\omega_n) \ , \quad g_{sy} = 0 \ ,\\
g_w^{p\eta} &=& -2i\sqrt{\frac{m_w}{2}}\frac{{\rm sign}(\bar{\omega}_n+\eta\varepsilon_I)}{\sqrt{\eta\varepsilon_0+\mu_w + \varepsilon_{so}+i\bar{\omega}_n}} \ .
\end{eqnarray}
Here $g_{k\eta} = (g_{k0} + \eta g_{ky})$, $\bar{g}_w^{p\eta} = -[g_w^{p(-\eta)}]^*$, and $\varepsilon_I = {\rm Im}(\varepsilon_0)$. Finally, we explicitly show the expressions for the normal Green's functions in the wire:
\begin{eqnarray}
g_{w0} &=& -i\sqrt{\frac{m_w}{2}}\sum_{\eta = \pm 1}\frac{(\varepsilon_0+\eta\varepsilon_{so}){\rm sign}(\bar{\omega}_n + \eta\varepsilon_I)}{\varepsilon_0\sqrt{\eta\varepsilon_0+\mu_w + \varepsilon_{so}+i\bar{\omega}_n}} \ ,\\
g_{wy} &=& ih\sqrt{\frac{m_w}{2}}\sum_{\eta = \pm 1}\frac{\eta\,{\rm sign}(\bar{\omega}_n + \eta\varepsilon_I)}{\varepsilon_0\sqrt{\eta\varepsilon_0+\mu_w + \varepsilon_{so}+i\bar{\omega}_n}} \ .
\end{eqnarray}

%
%
Note that in the absence of the spin-orbit coupling, zero magnetic field and for energy independent DOS in the wire
the self-consistency equation formally coincides with the one obtained in the seminal work by McMillan \cite{McMillan}.

\begin{figure*}
\centering
 \includegraphics[width=1\textwidth]{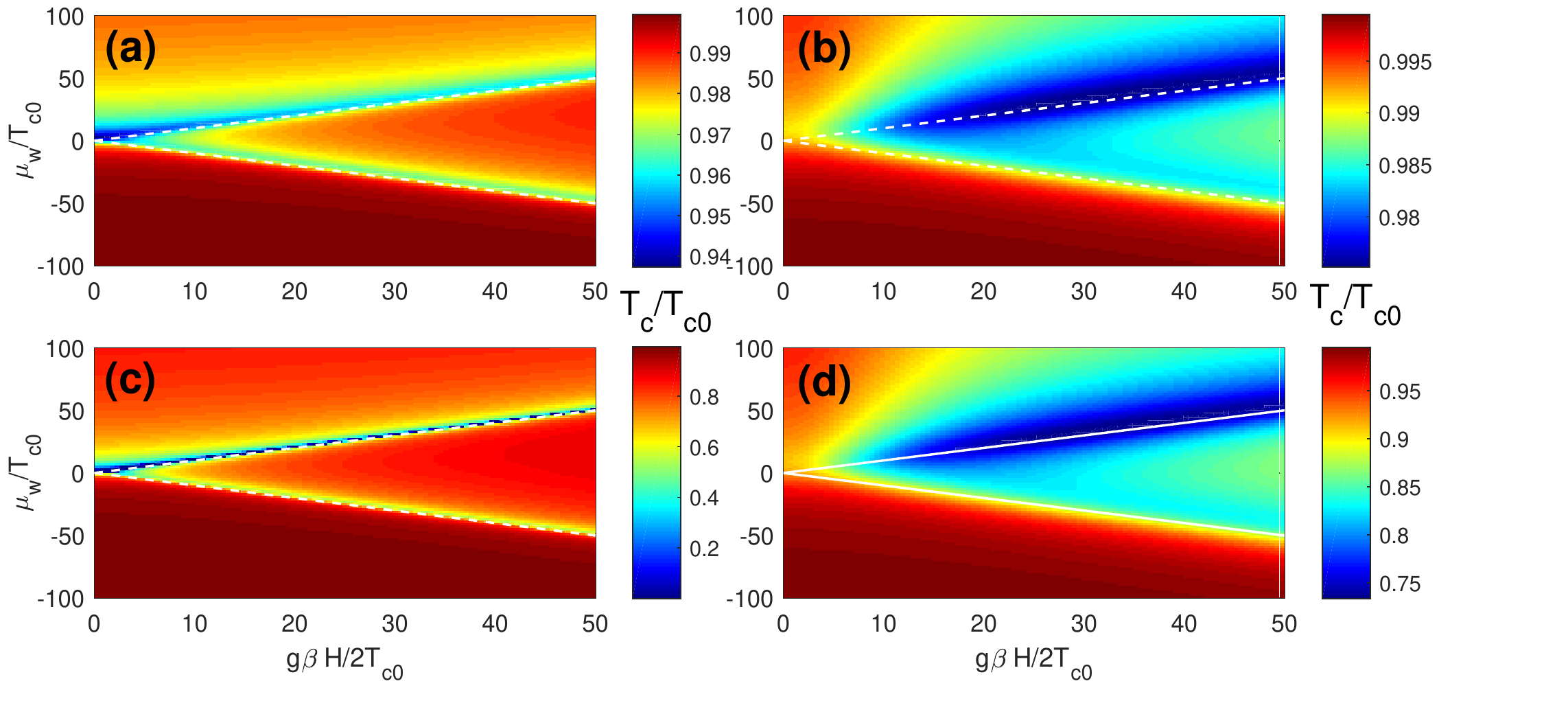}
 \caption{Color plot of the critical temperature of the system versus the chemical potential $\mu_w$ and the Zeeman energy $h=g\beta H/2$ for $\varepsilon_{so} = m_w\alpha ^2 = 0.46T_{c0}$ and several values of $\Gamma_s = t^2\ell_c R_w m_s\varphi_0/4\pi$ and $\Gamma_w = t^2\ell_c/2\ups_0$ with $\ups_0 = \sqrt{2T_{c0}/m_w}$.
 In panels (a) and (b) $\Gamma_w = 0.1 T_{c0}$, while in panels (c) and (d) we take $\Gamma_w = T_{c0}$.
 In panels (a) and (c) $\Gamma_s = 0.1 T_{c0}$, in panels (b) and (d) $\Gamma_s = 10 T_{c0}$.
 In all panels the white dashed lines denote the boundaries between nontopological and topological regimes $\mu_w=\pm h$.
 }\label{Fig:Tc_color_plot}
\end{figure*}

Turning now to the case of nonzero Zeeman energy and spin-orbit coupling we use numerical approach to analyze the solution of the self-consistency equation (\ref{SelfCons}) with the expression (\ref{anom_f_s0}~-~\ref{anom_f_sy}) for the anomalous Green function. Typical dependencies of the critical superconducting temperature
vs magnetic field and chemical potential $\mu_w$ are shown in Fig.~\ref{Fig:Tc_color_plot}.
Note that here we choose the strength of the spin-orbit coupling consistent with the properties of InAs \cite{Stanescu2011}:
$\varepsilon_{so} = m_w\alpha ^2 = 52\mu eV\simeq 600 mK$. Taking the critical temperature of Al $T_{c0}\simeq 1.3 K$ we find
$\varepsilon_{so} = m_w\alpha ^2 = 0.46T_{c0}$.

\begin{figure}
\centering
 \includegraphics[width=0.5\textwidth]{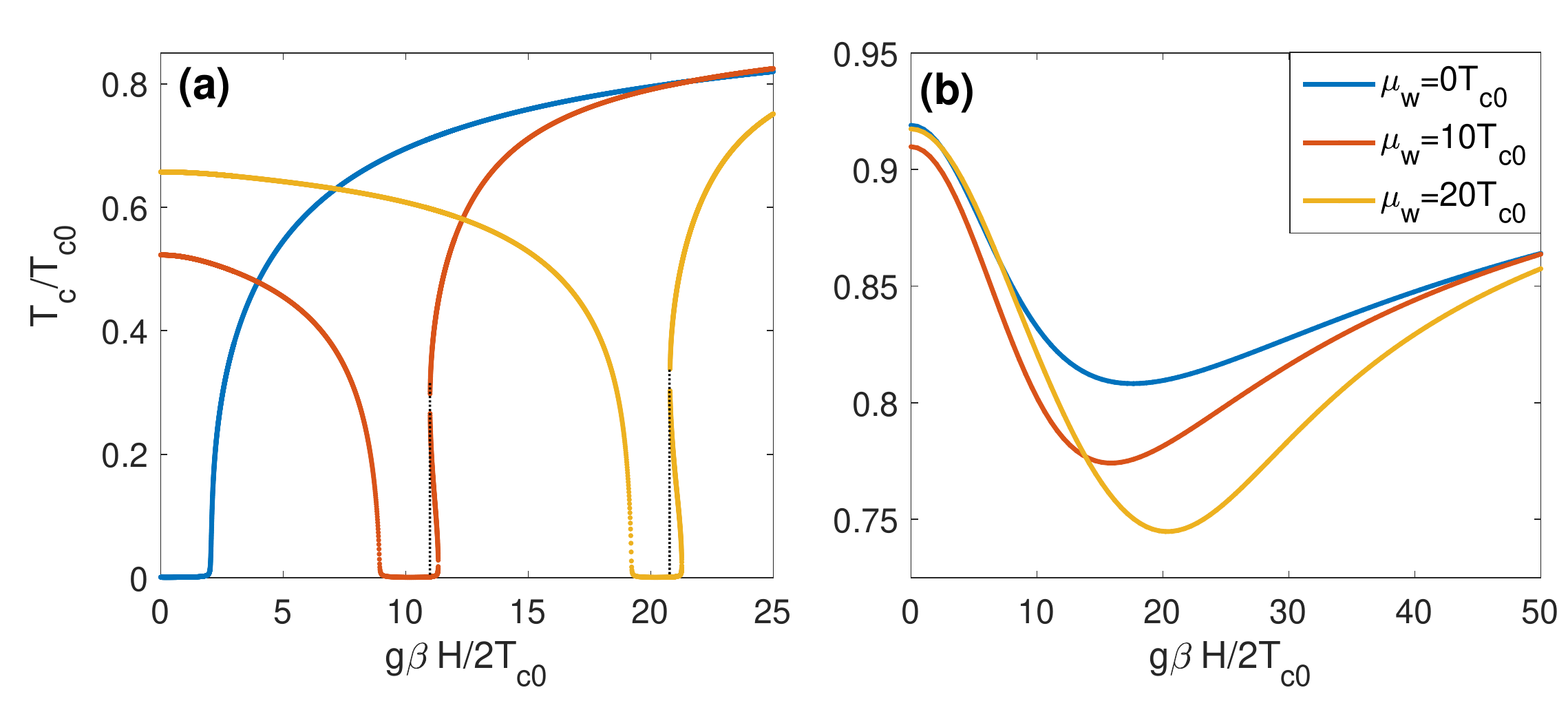}
 \caption{The critical temperature of the system as a function of the Zeeman field $h$ for
 different values of the chemical potential in the wire $\mu_w$ (shown in the legend). Here
 $\varepsilon_{so} = 0.46T_{c0}$ and $\Gamma_w = 1$. (a) $\Gamma_s = 0.1 T_{c0}$ and (b) $\Gamma_s = 10 T_{c0}$.}
 \label{Fig:Tc_cuts}
 \end{figure}

The color plots in Fig.~\ref{Fig:Tc_color_plot} show the critical temperature $T_c$ both in topologically trivial ($|\mu_w|>h$) and nontrivial ($|\mu_w|<h$) regimes.
The border lines $\mu_w=\pm h$ (shown by white dashed lines) coincide with the locations of van Hove singularities in the SM nanowire.
One can clearly see that the suppression of the critical temperature appears to be the strongest close to these lines.
The magnetic field dependence of $T_c$ appears to be drastically different in topologically trivial and nontrivial regimes.
Indeed, in nontopological regime the critical temperature decays as we increase the magnetic field due to a standard paramagnetic effect.
On the contrary, in topologically nontrivial regime $T_c$ increases (with or without initial decay at small fields).
This increase in the critical temperature originates from the reduction of the proximity effect due to almost pure spin polarization of
quasiparticles in the wire.
Of course the above mentioned increase in the critical temperature is limited from above by
either orbital or intrinsic paramagnetic effect in the S shell and continues up to the upper critical field in the superconductor.
One can see that the scattering rates $\Gamma_w$ and $\Gamma_s$ have a strong quantitative effect on the above physical picture because of the smearing and shifting of the peculiarities of DOS and resulting smoothing of $T_c$ variations.
The nonmonotonic behavior of $T_c$ is illustrated by the plots in Fig.~\ref{Fig:Tc_cuts}.
\rev{Using the above expressions~(\ref{G_s}~-~\ref{G_w}) and~(\ref{tun_rates_ratio}) for the tunneling rates,
we estimate the ratio of mode numbers as $N_w/N_s\sim 10^{-5}-10^{-4}$ for typical Majorana nanowires~\cite{Mourik2012,Chang2015,Higginbotham2015,Krogstrup2015,Albrecht2016,Leo}
and accounting of the decrease of $\ups_0$ value close to the van Hove singularity ($\ups_{F_s}/\ups_0\sim 10^{2}-10^3$),
we get $\Gamma_w/\Gamma_s \sim 10^{-3}-10^{-1}$.
Assuming strong coupling between the nanowire and superconducting shell with $\Gamma_s\gtrsim T_{c0}$,
we get $\Gamma_w\sim(10^{-3}-10^{-1})T_{c0}$.
Note that in realistic experimental conditions the number of modes in the wire ($N_w$) can increase
due to the formation of the accumulation layer near the superconductor-semiconductor interface~\cite{Antipov2018,Mikkelsen2018,Woods2018}.
However, the increase of the shell thickness $d_s$ may weaken the effect in the multimode regime of the superconductor.
Overall, such estimate allows us to expect that the consequences of the inverse proximity effect analyzed in our paper can be observed experimentally.}

\begin{figure}
\centering
 \includegraphics[width = 0.4\textwidth]{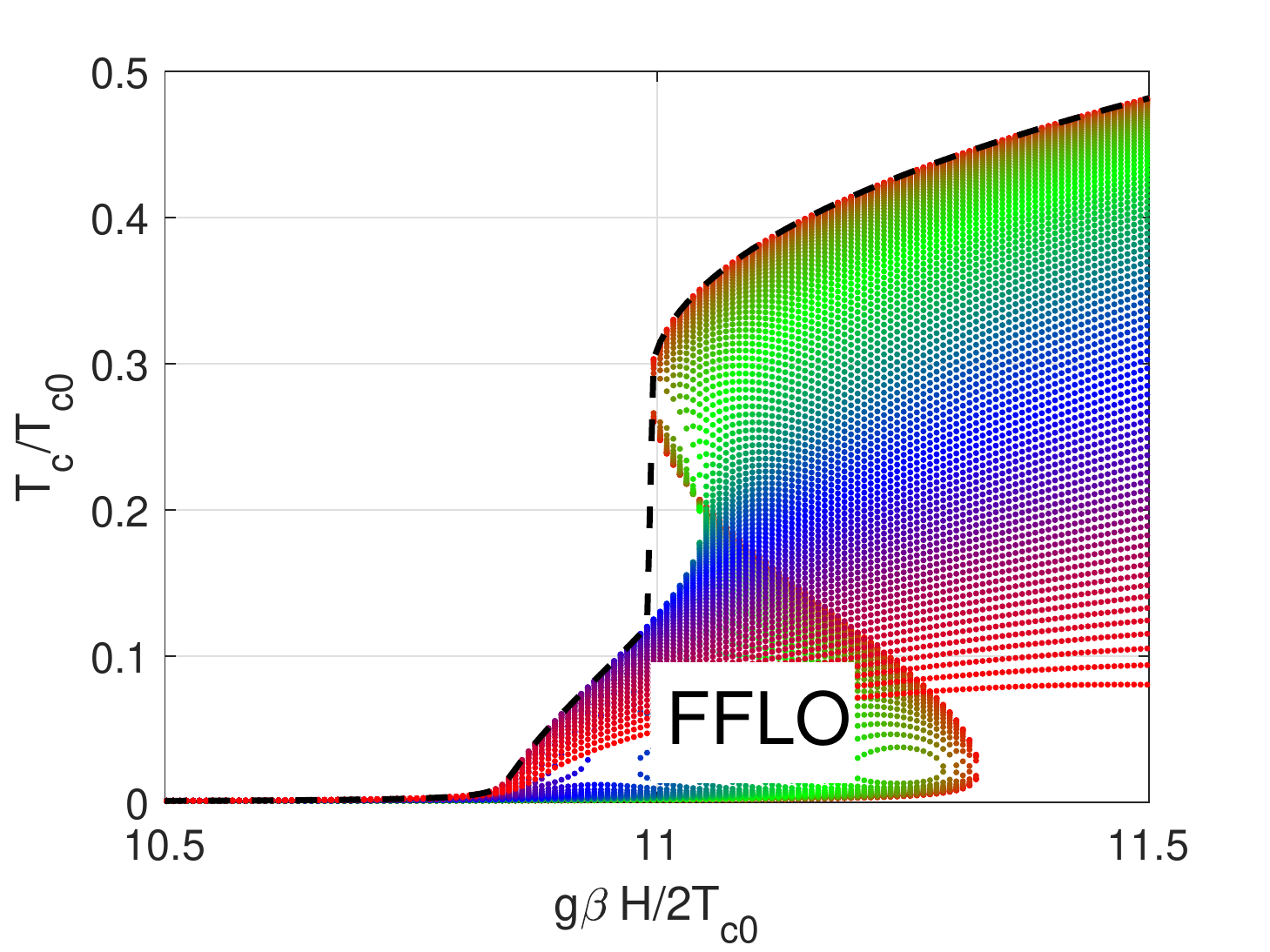}
 \caption{Critical temperature of the system as a function of the Zeeman field $h$ for $\varepsilon_{so} = 0.46T_{c0}$,
 $\Gamma_s \to 0$ and $\Gamma_w =1$ for the superconducting states with different modulation vectors q spreading from $q = 0.44 m_sT_{c0}/k_{Fs}$ at $h = 11.05 T_{c0}$ to $q = m_sT_{c0}/k_{Fs}$ at $h = 10.8 T_{c0}$. }
 \label{Fig:FFLO}
 \end{figure}

It is worth noting that the $T_c(h)$ plot in the panel (a) of the Fig.~\ref{Fig:Tc_cuts} clearly demonstrates the appearance of the $h$ regions where the linearized
self-consistency equation has three solutions instead of one. In other words, there can exist three critical temperatures corresponding to a given magnetic field.
\rev{This is the evidence of that although the superconducting shell has a small g-factor, indirectly superconducting region is affected by effective Zeeman field through tunneling.}
The presence of several solutions for $T_c$ is typical for the standard paramagnetic effect in superconductors and usually this behavior results
in the FFLO instability of the homogeneous solution for the gap function \cite{SaintJames}.
To verify this scenario in our system we
have solved a self-consistency equation for the modulated order parameter $\Delta_s\propto e^{iqy}$ and found that the regions with several solutions
for $T_c$ for the homogeneous gap indeed can host more energetically favorable inhomogeneous FFLO gap function.
The critical temperature $T_c (q)$ for different $q$
values can be seen in Fig.~\ref{Fig:FFLO}. As we increase the $h$ value from $h = 10.8$ to $h = 11.05$ the $q$ value corresponding to the maximal $T_c$
changes from $k_{Fs}q/m_s=T_{c0}$ to $k_{Fs}q/m_s=0.44T_{c0}$.
\rev{It is important to note that as we solve the linearized equation for the superconducting gap, we find, of course, only the critical temperatures corresponding to the second-order phase transitions. Changing the period of the gap modulation of the FFLO type we also find only the temperatures corresponding to the second-order phase transition.
The physical picture can become more complicated if one takes into account possible first-order transitions corresponding to the interplay between different local minima of the thermodynamic potential in the nonlinear regime.
However, the solution of nonlinear gap equations is beyond the scope of the current work and needs further investigations.
Note also that the possible FFLO phase appears on either side of topological transition ($h^2=\mu_w^2+\Delta_{ind}^2$) depending on the sign of the chemical potential $\mu_w$. Indeed, in general the temperature as a formal solution of the self-consistency equation \eqref{SelfCons} is not a single-valued function of magnetic field in the regions $h\gtrsim \pm\mu_w$ slightly above the positions of van Hove singularities, being inside the topological (trivial) regime for the upper (lower) sign. In practical cases the upper in $h$ singularity is more prominent (see, e.g., Fig.~\ref{Fig:Tc_cuts}).
Additionally,}
an accurate analysis of FFLO state should include careful consideration of the modulation of the superconducting order parameter both along the wire and in the azimuthal direction~\cite{Samokhvalov2009,Samokhvalov2017}.

Before we conclude we discuss briefly the influence of the inverse proximity effect on the effective induced gap operator $\Delta_{top}$ in the topological regime \rev{$h^2>\mu_w^2+\Delta_{ind}^2$} which is of crucial importance for topological superconducting electronics and topologically protected fault-tolerant quantum computing.
In our estimates we take the standard limit of $\mu_w = 0$ for the sake of simplicity.
First, the increase of $\Gamma_w$ reduces
the parameter range of the topological insulator regime $\alpha p \gg h$ as the magnetic field should well exceed $\Gamma_w$ to avoid the suppression of the critical temperature due to the van Hove singularities (see Fig.~\ref{Fig:Tc_color_plot}(a) and (c) for small $\Gamma_s$ values).
As soon as $\Gamma_w$ becomes comparable with $\alpha p$ with the typical quasiparticle momentum $p\simeq \sqrt{2 m_w h}$
this regime completely disappears.
Further increase of the scattering rate should suppress the gap $\Delta_{top} \propto \alpha p \Delta_{ind}/h$ in the Kitaev limit.
Indeed, for $\Gamma_w>m_w\alpha^2 = \epsilon_{so}$ its value is limited from the above by
the quantity $\Delta_{top}\simeq\Delta_{ind} (\epsilon_{so}/\Gamma_w)^{1/2}<\Delta_{ind}$.
Such decrease in the attainable induced gap values imposes more strict conditions on working temperatures for Majorana-based devices,
due to quasiparticle poisoning as the residual quasiparticle density is exponentially sensitive to the gap values (see, e.g., \cite{Saira_nqp_2012,Knowles2012,Maisi2013,Woerkom2015}).
Of course, at large $\Gamma_s$ values (see Fig.~\ref{Fig:Tc_color_plot}(b) and (d)) the van Hove singularities are smeared and the critical temperature (together with the gap value) is suppressed only partially.
However, even the partial suppression of a few to tens \% may drastically increase the effect of quasiparticle poisoning mentioned above.

\section{Conclusion}
To sum up, we have studied the distinctive features of the inverse proximity effect arising in the presence of a large Zeeman energy and strong spin-orbit coupling in the hybrid systems consisting of the SM nanowires covered by thin superconducting films. Assuming a strong difference in $g$-factors between the wire and superconducting metal we find the range of parameters and fields corresponding to the FFLO instability and the regime of reentrant superconductivity.
We focus on the topologically nontrivial regime of relatively large magnetic fields and analyze consequences of the inverse proximity effect on
the quasiparticle poisoning in Majorana-based devices.


\begin{acknowledgements}
We are pleased to thank J.~P.~Pekola for valuable and stimulating discussions.
This work has been supported in part
by the Russian Foundation for Basic Research (A.~A.~K.),
 German Research Foundation (DFG) Grant No. KH 425/1-1 (I.~M.~K.),
by the Russian Science Foundation, Grant No. 17-12-01383 (A.~S.~M.),
Foundation for the advancement of theoretical physics ``BASIS'',
\rev{and the Academy of Finland Grant No. 298451.} 
\end{acknowledgements}

\appendix
\section{Appendix: Derivation of the model equations}\label{App:Model_derivation}
Here we present the derivation of the model equations~(\ref{gorkov_eqs_s}~-~\ref{gorkov_eqs_w}).
The Hamiltonian of the system \eqref{hamiltonian} given in the main text consists of three terms (\ref{hamiltonian_s}~-~\ref{hamiltonian_t}) describing the superconducting shell, the semiconducting nanowire, and the tunneling terms between these subsystems, respectively.
The notations are also given in the main text.

The field operators in the shell are normalized in the following way:
 \begin{equation}
 \left[ \psi_{\sigma}(\mathbf{r}), \psi_{\sigma'}^{\dagger}(\mathbf{r}')\right]_+ = \frac{1}{d_sR_w}\delta_{\sigma \sigma'} \delta(\varphi-\varphi')\delta(y-y').
 \end{equation}
Here $[A,B]_+=AB+BA$ denotes the anticommutator of two operators $A$ and $B$.
The operators in the nanowire satisfy the following anticommutation relations:
 \begin{equation}
 \left[a_{\sigma}(y), a_{\sigma'}^{\dagger}(y')\right]_+ = \frac{1}{S_w}\delta_{\sigma \sigma}\delta(y-y') \ .
 \end{equation}

Now we proceed with the definitions of the Matsubara Green's functions in the shell. We adopt the following spinor notation in the Nambu space:
\begin{equation}\label{spinor_notation}
\Psi(\mathbf{x}) = \begin{pmatrix}\psi_{\uparrow}(\mathbf{x})\\ \psi_{\downarrow}(\mathbf{x})\end{pmatrix} \ \text{and} \ \widetilde{\Psi}(\mathbf{x}) = \left(\psi_{\uparrow}(\mathbf{x}) \ \psi_{\downarrow}(\mathbf{x}) \right) \ .
\end{equation}
Here $\mathbf{x} = (\mathbf{r},\tau)$, while $\tau$ is the imaginary time variable in the standard Matsubara technique. Using the above notations, we introduce the Green's functions as follows:
\begin{gather}\label{GF_definitions}
\check{G}_s = \left\langle T_{\tau}\left[\begin{pmatrix}-\Psi(\mathbf{x}_1)\\ \widetilde{\Psi}^{\dagger}(\mathbf{x}_1)\end{pmatrix}\otimes \left(\Psi^{\dagger}(\mathbf{x}_2) \ -\widetilde{\Psi}(\mathbf{x}_2)\right) \right]\right\rangle =
\begin{pmatrix}\hat{G}_s(\mathbf{x}_1,\mathbf{x}_2)&\hat{F}_s(\mathbf{x}_1,\mathbf{x}_2)\\ \hat{F}_s^{\dagger}(\mathbf{x}_1,\mathbf{x}_2)&\hat{\bar{G}}_s(\mathbf{x}_1,\mathbf{x}_2)\end{pmatrix} \ ,
\end{gather}
where $\langle...\rangle$ denotes the Gibbs statistical average and $T_{\tau}$ is the time-ordering operator. The definition of the Green's functions of the wire is given by Eqs.~(\ref{spinor_notation}) and~(\ref{GF_definitions}) with the replacement of the field operators $\psi_{\alpha}(\mathbf{x})\to a_{\alpha}(\mathbf{y})$ along with the replacement of the subscripts $s\to w$, where $\mathbf{y} = (y,\tau)$.

Using the Nambu spinor notation presented in the above equation, we introduce the mixed Green's functions in the spin-Nambu space:
\begin{eqnarray}\label{mixed_GF_definitions}
\check{G}_t = \left\langle T_{\tau}\left[\begin{pmatrix}-\Psi(\mathbf{x}_1)\\ \widetilde{\Psi}^{\dagger}(\mathbf{x}_1)\end{pmatrix}\otimes \left(a^{\dagger}(\mathbf{y}_2) \ -\widetilde{a}(\mathbf{y}_2)\right) \right]\right\rangle =
\begin{pmatrix}\hat{G}_t(\mathbf{x}_1,\mathbf{y}_2)&\hat{F}_t(\mathbf{x}_1,\mathbf{y}_2)\\ \hat{F}_t^{\dagger}(\mathbf{x}_1,\mathbf{y}_2)&\hat{\overline{G}}_t(\mathbf{x}_1,\mathbf{y}_2)\end{pmatrix} \ .
\end{eqnarray}
We start the derivation of Gor'kov equations~(\ref{gorkov_eqs_s}~-~\ref{gorkov_eqs_w}) by writing the equations for the mixed Matsubara Green's functions~(\ref{mixed_GF_definitions}) in the frequency-coordinate representation:
\begin{eqnarray}
\label{Wire_Direct_G_s}
[\check{G}_{s}^{(0)}(\mathbf{r}_1)]^{-1}\check{G}_t(\mathbf{r}_1,y_2)-(1/\zeta)\check{\mathcal{T}}(\mathbf{r}_1)\check{G}_w(y_1,y_2) = 0 \ ,\\
\label{Wire_Direct_G_w}
\check{G}_t(\mathbf{r}_1,y_2)[\check{G}_{w}^{(0)}(y_2)]^{-1} - \zeta \left\langle \check{G}_s(\mathbf{r}_1,\mathbf{r}_2)\check{\mathcal{T}}(\mathbf{r}_2)\right\rangle_{\varphi_2} = 0 \ ,
\end{eqnarray}
where
\begin{eqnarray}
[\check{G}_{s}^{(0)}(\mathbf{r})]^{-1} &=& i\omega_n-\varepsilon_s(\mathbf{r})\check{\tau}_z+\check{\Delta} \ , \\
{}[\check{G}_{w}^{(0)}(y)]^{-1} &=& i\omega_n-\varepsilon_{w}(y)+i\alpha  \partial_y \hat{\sigma}_x - h\hat{\sigma}_y \ ,
\end{eqnarray}
$\zeta = \sqrt{d_s/\pi R_w}$, $\check{\mathcal{T}}(\mathbf{r}) =[\mathcal{T}(\mathbf{r})\check{\Pi}_{z_+}-\mathcal{T}^{\dagger}(\mathbf{r})\check{\Pi}_{z_-}]$, $\mathcal{T}(\mathbf{r})$ is the tunneling amplitude, $\check{\Pi}_{z_{\pm}} = (1\pm\check{\tau}_z)/2$, and $\langle ... \rangle_{\varphi} = \int d\varphi$.
The solution of Eq.~\eqref{Wire_Direct_G_s} with the boundary conditions corresponding to the isolated superconductor
takes the form
\begin{equation}\label{mixed_solution}
\check{G}_t(\mathbf{r}_1,y_2) = \sqrt{d_sR_wS_w} 
\langle\check{G}_s^{(0)}(\mathbf{r}_1,\mathbf{r})\check{\mathcal{T}}(\mathbf{r})\check{G}_w(y,y_2)\rangle_{\mathbf{r}} \ ,
\end{equation}
where $\langle ... \rangle_{\mathbf{r}} = \int dy \ d\varphi$. The Green's function of the isolated superconductor satisfies the following equation:
\begin{gather}
[\check{G}_{s}^{(0)}(\mathbf{r}_1)]^{-1}\check{G}_s^{(0)}(\mathbf{r}_1,\mathbf{r}_2) =
\check{1}(d_sR_w)^{-1}\delta(\varphi_1-\varphi_2)\delta(y_1-y_2) \ .
\end{gather}

Substituting the Eq.~(\ref{mixed_solution}) into the equation for the Green's function in the wire
\begin{gather}\label{gorkov_wire_app}
\check{G}_{w}^{-1}(y_1)\check{G}_{w}(y_1,y_2)-\zeta\langle \check{\mathcal{T}}^{\dagger}(\mathbf{r}_1)\check{G}_t(\mathbf{r}_1,y_2) \rangle_{\varphi_1} =
\check{1}S_w^{-1}\delta(y_1-y_2) \
\end{gather}
and performing the ensemble averaging over the random tunneling amplitudes
\begin{equation}
\overline{\mathcal{T}(\mathbf{r})\mathcal{T}(\mathbf{r}')} =  t^2\ell_c\delta(y-y')\delta(\varphi-\varphi') \ ,
\end{equation}
we get Dyson-Gor'kov equations for the Green's functions of the wire
\begin{gather}\label{gorkov_wire_app_pt}
\left[\check{G}_w^{-1}(y_1)-\check{\Sigma}_w(y_1)\right]\check{G}_w(y_1,y_2) =
\check{1}S_w^{-1}\delta(y_1-y_2) \ ,
\end{gather}
with the self-energy part taken in the limit of an isolated superconducting shell
\begin{equation}\label{gorkov_wire_Self_Energy}
\check{\Sigma}_w(y) = d_s R_w t^2\ell_c\check{\tau}_z\langle\check{G}_s^{(0)}(\mathbf{r},\mathbf{r})\rangle_{\varphi}\check{\tau}_z \ .
\end{equation}

The Dyson-Gor'kov equations in the superconducting shell are derived using the same arguments as for the previous case
\begin{gather}\label{gorkov_shell_app_int}
\left[\check{G}_s^{-1}(\mathbf{r}_1)-\check{\Sigma}_s(\mathbf{r}_1)\right]\check{G}_s^{(0)}(\mathbf{r}_1,\mathbf{r}_2) =
\check{1}(d_sR_w)^{-1}\delta(\varphi_1-\varphi_2)\delta(y_1-y_2) \ ,
\end{gather}
with the self-energy part taken in the limit of an isolated semiconducting wire
\begin{equation}\label{gorkov_shell_Self_Energy}
\check{\Sigma}_s(\mathbf{r}) = S_wt^2\ell_c\check{\tau}_z\check{G}^{(0)}_w(y,y)\check{\tau}_z \ ,
\end{equation}
the solution of Eq.~\eqref{Wire_Direct_G_w}, and with the Green's function of the isolated wire satisfying the following equation
\begin{equation}\label{gorkov_wire_app_nonint}
\check{G}_{w}^{-1}(y_1)\check{G}^{(0)}_{w}(y_1,y_2) = \check{1}S_w^{-1}\delta(y_1-y_2) \ .
\end{equation}
To take exact boundary conditions into account we follow the procedure suggested in \cite{McMillan}
replacing $\check{G}^{(0)}_{s}(\mathbf{r}_1,\mathbf{r}_2)$ and $\check{G}^{(0)}_{w}(y_1,y_2)$ in Eqs.~(\ref{gorkov_wire_Self_Energy}) and~(\ref{gorkov_shell_Self_Energy}) with the exact ones
\begin{eqnarray}
\label{gorkov_Self_Energy_w}
\check{\Sigma}_w(y) &= d_sR_wt^2\ell_c\check{\tau}_z\langle\check{G}_s(\mathbf{r},\mathbf{r})\rangle_{\varphi}\check{\tau}_z \ ,\\
\label{gorkov_Self_Energy_s}
\check{\Sigma}_s(\mathbf{r}) &= S_wt^2\ell_c\check{\tau}_z\check{G}_w(y,y)\check{\tau}_z \ .
\end{eqnarray}

Finally the Fourier transform of Eqs.~\eqref{gorkov_wire_app_pt} and \eqref{gorkov_shell_app_int} with the self-energies \eqref{gorkov_Self_Energy_w} and \eqref{gorkov_Self_Energy_s}~\rev{along with the renormalization of the Green's functions $\check{G}_w\to\check{G}_w/S_w$ and $\check{G}_s\to\check{G}_s/d_s$} completes the derivation of Eqs.~(\ref{gorkov_eqs_s}~-~\ref{gorkov_eqs_w}) in the main text.

\vspace{3cm}

\end{document}